# Solving the Black Box Problem:
# A Normative Framework for Explainable Artificial Intelligence

Carlos Zednik
Otto-von-Guericke-Universität Magdeburg
carlos.zednik@ovgu.de
http://carloszednik.net/

**Abstract**

Many of the computing systems programmed using Machine Learning are opaque: it is difficult to know why they do what they do or how they work. The Explainable Artificial Intelligence research program aims to develop analytic techniques with which to render opaque computing systems transparent, but lacks a normative framework with which to evaluate these techniques' explanatory success. The aim of the present discussion is to develop such a framework, while paying particular attention to different stakeholders' distinct explanatory requirements. Building on an analysis of 'opacity' from philosophy of science, this framework is modeled after David Marr's influential account of explanation in cognitive science. Thus, the framework distinguishes between the different questions that might be asked about an opaque computing system, and specifies the general way in which these questions should be answered. By applying this normative framework to current techniques such as input heatmapping, feature-detector identification, and diagnostic classification, it will be possible to determine whether and to what extent the Black Box Problem can be solved.

## 1. Introduction

Computing systems programmed using Machine Learning (ML) are increasingly capable of solving complex problems in Artificial Intelligence (AI). Unfortunately, these systems remain characteristically *opaque*: it is difficult to "look inside" so as to understand why they do what they do or how they work.

Opacity is the heart of the *Black Box Problem*—a problem with significant practical, legal, and theoretical consequences. Practically, end-users are less likely to trust and cede control to machines whose workings they do not understand (Burrell, 2016; Ribeiro, Singh, & Guestrin, 2016), and software engineers may be unable to intervene in order to quickly and systematically improve performance (Hohman, Kahng, Pienta, & Chau, 2018). Legally, opacity prevents regulatory bodies from determining whether a particular system processes data fairly and securely (Rieder & Simon, 2017), and may hinder end-users from exercising their rights under the European Union's *General Data Protection Regulation* (European Commission, 2016). Theoretically, the Black Box Problem makes it difficult to evaluate the potential similarity between artificial neural networks and biological brains (Buckner, 2018), and to determine the extent to which computers being developed using ML may be considered genuinely intelligent (Zednik, 2018).

Investigators within the *Explainable AI* (XAI) research program intend to ward off these consequences through the use of analytic techniques with which to render opaque computing systems *transparent*.[1] Although the XAI research program has already commanded significant attention (Burrell, 2016; Doran, Schulz, & Besold, 2017; Lipton, 2016; Ras, van Gerven, & Haselager, 2018; Zerilli, Knott, Maclaurin, & Gavaghan, 2018), important normative questions remain unanswered. Most fundamentally, it remains unclear how Explainable AI *should* explain: what is required to render opaque computing systems transparent? Given different stakeholders' distinct reasons for interacting with such computing systems, these stakeholders will require different kinds of explanations. Moreover, it remains unclear how explanation in this context relates to other epistemic achievements such as description, prediction, intervention, and understanding—each of which might be more or less important in different domains, for different stakeholders, and with respect to different systems. Finally, it remains unclear what the prospects are of actually explaining the behavior of ML-programmed computing systems as they become increasingly powerful, sophisticated, and widespread.

The present discussion contributes to the maturation of the Explainable AI research program by developing a normative framework for rendering opaque computing systems transparent. This framework accommodates different stakeholders' distinct explanatory requirements, and specifies the questions that should be answered in order to ward off the Black Box Problems practical and theoretical consequences. Notably, inspiration will be sought in cognitive science. Indeed, insofar as the problem of rendering opaque computing systems transparent is not unlike the problem of explaining the behavior of humans and other biological cognizers, the Explainable AI research program can benefit from co-opting some of the norms and practices of cognitive science (see also Rahwan et al., 2019).

The discussion proceeds as follows. Section 2 builds on previous philosophical work by Paul Humphreys (2009) to analyze the oft-used but nevertheless ill-understood notions of 'opacity' and 'transparency'. Section 3 then invokes Tomsett et al.'s (2018) notion of an 'ML ecosystem' to distinguish between the stakeholders who are most likely to seek explanations of a particular system's behavior. It also introduces David Marr's (1982) *levels of analysis* account of explanation in cognitive science so as to better understand these stakeholders' distinct explanatory requirements. Indeed, different stakeholders in the ML ecosystem can be aligned with different questions in Marr's account—questions about *what*, *why*, *how*, and *where* a computer program is carried out. Sections 4 and 5 then introduce and evaluate the explanatory contributions of several current analytic techniques from Explainable AI: *layer-wise relevance propagation* (Montavon, Samek, & Müller, 2018), *local interpretable model-agnostic explanation* (Ribeiro et al., 2016), *feature-detector* identification *(Bau et al., 2018)*, and *diagnostic classification* (Hupkes, Veldhoen, & Zuidema, 2018).[2] Indeed, these sections show that different XAI techniques are capable of answering distinct questions, and thus, are likely to satisfy different stakeholders' explanatory requirements.

---

1   There are two distinct streams within the Explainable AI research program. The present discussion focuses on attempts to *solve* the Black Box Problem by analyzing computing systems so as to render them transparent *post hoc,* i.e., after they have been developed or deployed. In contrast, the discussion will not consider efforts to *avoid* the Black Box Problem altogether, by modifying the relevant ML methods so that the computers being programmed do not become opaque in the first place (for discussion see, e.g., Doran, Schulz, & Besold, 2017).

## 2. The Black Box Problem in Artificial Intelligence

*2.1 From Machine Learning to the Black Box Problem*

The Black Box Problem is traditionally said to arise when the computing systems that are used used to solve problems in AI are *opaque*. This manner of speaking is grounded in the metaphorical intuition that a system's behavior can be explained by "looking inside" so as to understand why it does what it does or how it works. Although most computing systems are constructed from well-understood hardware components that afford no literal obstacle to "looking inside", they might nevertheless be considered opaque in the sense that it is difficult to know exactly how they are programmed.

Machine Learning is just one approach among many for programming computers that solve complex problems in AI. Unlike their colleagues working within other AI approaches, however, developers in Machine Learning exert limited influence on the way in which the relevant problems are solved. Of course, ML developers must decide on basic architectural principles such as whether the system takes the form of a deep neural network, a support vector machine, a decision tree, or some other kind of system with a particular set of learnable parameters. Moreover, they must choose an appropriate learning algorithm, and must identify a suitable learning environment, in which the learnable parameters can obtain values with which to solve the problem at hand. Nevertheless, ML developers do not typically decide on the particular values these parameters eventually obtain (e.g. the weights of individual network connections), and in this sense, do not wholly determine the way in which the problem is actually solved.

This relative lack of influence is a great advantage insofar as Machine Learning methods are often capable of identifying highly unintuitive and subtle solutions that are unlikely to be found using more traditional methods. Indeed, it is this capability that explains the recent influx of Machine Learning in many different domains, as well as its great promise for society as a whole.[3] Unfortunately, however, this great advantage also comes at a significant cost. Unlike the computing systems being programmed using more traditional methods, the systems being programmed using Machine Learning are characteristically opaque.

---

2  This is by no means a comprehensive list of techniques, and inclusion in this list should not be taken to indicate that these techniques are superior to others. Indeed, XAI is an incredibly dynamic field in which original and increasingly powerful techniques are being developed almost on a daily basis. But although not all relevant XAI techniques can be considered here, the normative framework being developed is meant to apply generally, including to those techniques that have not been considered or that may not have yet been developed.

3  Domains in which ML-programmed systems have recently had a great, if not revolutionary, influence include game-playing, autonomous driving and flying, question-answering, natural language processing, machine vision, behavior-prediction, and product-recommendation, among many others. That said, successes within these domains should not, of course, be taken to imply that Machine Learning methods are all-conquering. Indeed, many important AI problems remain unsolved, and in many cases, it is unclear whether, and if so how, ML methods could ever be used to solve them. Indeed, in many problem domains, traditional AI methods remain far more effective than the methods developed in Machine Learning (for discussion see, e.g., Lake, Ullman, Tenenbaum, & Gershman, 2017;

On the face of it, the reasons for this characteristic opacity are obvious. Because developers exert relatively limited influence on the setting of parameter values, they might not know how these parameters contribute to the system's behavior. In addition, even if individual parameter values are known, the fact that they might interact nonlinearly as well as recurrently means that it is almost impossible to understand, predict, or systematically intervene on the way in which a particular input is transformed so as to generate a particular output.

Upon closer inspection, however, it is clear that a system's opacity cannot be reduced to an ML developers' knowledge of parameter values. For example, many deep neural networks have been found to learn high-level representations that capture abstract properties of the learning environment, but that do not map neatly onto individual parameter values (see, e.g., Bau et al., 2018; Buckner, 2018). In such cases, knowledge of the relevant representations is arguably far more relevant for the purposes of explaining the system's behavior than knowledge of individual connection weights.[4] In addition, it is not clear that ML developers are the only stakeholders with respect to which to assess a particular system's opacity. Many other stakeholders, from end-users to regulatory bodies, are likely to seek a better understand of a particular system's behavior, despite being unable or unwilling to acquire knowledge of the system's learnable parameters. Thus, although it may be tempting to reduce a system's opacity to a developer's knowledge of learnable parameters, opacity is in fact a highly nuanced phenomenon that bears clarification. What is required, in other words, is a better understanding of the dual notions of 'opacity' and 'transparency', and thus, a better understanding of the Black Box Problem itself.

*2.2 What is the Black Box Problem?*

One of the most influential recent analyses of 'opacity' is due to Paul Humphreys (2009). Although Humphreys' work is primarily concerned with computer simulations in scientific disciplines such as high-energy physics and molecular biology, his analysis of 'opacity' is also a useful starting point for discussions of the Black Box Problem in Artificial Intelligence. On Humphreys' analysis, computing systems are

> "opaque relative to a cognitive agent $X$ at time $t$ just in case $X$ does not know at $t$ all of the epistemically relevant elements of the [system]" (Humphreys, 2009, p. 618).[5]

Marcus, 2018)
4   Indeed, ML developers typically do have access to the values of learnable parameters. Nevertheless, these developers are often the first to call for additional explanations so as to, e.g., demonstrate that a system does in fact do what it is supposed to do (see Section 4), or to improve its performance when it does not (See Section 5 and Hohman, Kahng, Pienta, & Chau, 2018).
5   Humphreys subsequently introduces the notion of *essential* epistemic opacity, which applies to systems whose epistemically relevant elements are not only unknown to the agent, but that are in fact *impossible* to know by that agent (Humphreys 2009, p. 618). Notably, Humphreys' notion of possibility is not logical, but practical—it depends on an agent's limited time, money, and computational power,

Two features of this analysis are worth emphasizing. First, opacity is *agent-relative*. That is, a computing system is never opaque in and of itself, but opaque only with respect to some particular agent. Second, opacity is an *epistemic* property: it concerns the agent's (lack of) knowledge. According to Humphreys, this knowledge concerns the system's *epistemically relevant elements* (EREs). Although Humphreys does not further analyze the notion of an epistemically relevant element, it can be fleshed out in a way that suits current purposes. Thus, an *element* can be understood as, for example, a step in the process of transforming inputs to outputs, or as a momentary state-transition within the system's overall evolution over time. An *epistemically relevant* element is one which is not only known to the agent, but which can be cited by him or her to explain the occurrence of some other element, or of the system's overall output.

The term 'explain' is left intentionally ambiguous. A computing system's epistemically relevant elements may take the form of physical structures, mathematical states-of-affairs, or even reasons—among many other things. Accordingly, the presence or absence of any such element might be explained physically (e.g. by appealing to the magnetization of hardware registers), mathematically (e.g. by appealing to binary strings), or even rationally (e.g. by appealing to the possibility that the relevant binary string represents a particular goal state). In general, different kinds of explanations invoke different kinds of EREs, and different kinds of EREs are likely to be appropriate to different agents and to different systems.

Two notes on the ambiguity of 'explain'. First, for any given system, many different EREs can be considered equally "real", and thus, many different explanations can be considered equally legitimate. By way of analogy to the computing system just above, the behavior of humans and other biological cognizers can also be explained physically (e.g. by reference to neurobiological mechanisms), mathematically (e.g. by reference to computational processes), and rationally (e.g. by reference to beliefs and desires). Staunchly reductionist or eliminativist philosophical tendencies notwithstanding (e.g., Bickle, 2006; Churchland, 1981), all of these explanations are equally legitimate, and the EREs invoked therein are equally "real". The situation in Explainable AI is analogous: many different kinds of explanations can and should be considered legitimate for the purposes of rendering opaque computing systems transparent.

But although many different kinds of explanations are equally legitimate, not all of them are likely to be equally useful for any individual agent. Whereas some agents may in fact seek to explain a computing system's behavior by invoking as EREs the values of learnable parameters, many other agents—in particular, agents with more limited cognitive resources, with a more limited understanding of computer programming, or with altogether different goals and interests—will require rather different explanations.

among other resources (see, e.g., Durán & Formanek, 2018). For this reason, analytic techniques from XAI that allow an agent to use these resources more efficiently are poised to reduce the scope of the impossible, all while allowing that some systems may remain opaque even after the most powerful techniques have been applied. That said, the present discussion need not speculate about the scope of the impossible. The primary aim of this discussion is to show how systems that *can* be rendered transparent *should* be rendered transparent.

Thus, whereas some agents might be satisfied by explanations that cite a system's parameter values, others are likely to prefer explanations that cite higher-order representations. Moreover, some agents may require explanations that invoke the physical properties of hardware components, whereas others may instead invoke explanations that cite environmental features such as objects, colors, persons, or intentions. Thus, the ambiguity in the term 'explain' shows why the notions of 'opacity' and 'transparency' cannot be properly understood solely by considering developers' (lack of) knowledge of learnable parameters: For many agents, learnable parameters are simply not the right kind of ERE for the purposes of rendering opaque computing systems transparent.

The discussion so far shows that the sense in which computing systems are opaque—and thus, the sense in which they should be rendered transparent—differs between agents. At the same time, the outline of a general solution to the Black Box Problem is already beginning to taking shape: In order to render an opaque computing system transparent, a particular agent must seek knowledge of an appropriate set of epistemically relevant elements with which to explain that system's behavior. Two questions must be answered before this outline can be fleshed out into a normative framework for Explainable AI. Which agents are concerned with explaining the behavior of computers programmed using Machine Learning? What are, for these agents, the appropriate epistemically relevant elements? By answering these two questions, it will eventually be possible to determine how Explainable AI should explain, and to evaluate the explanatory success of current and future XAI analytic techniques.

**3. From Machine Learning to Marr**

*3.1. What are the stakeholders?*

Tomsett et al. (2018) provide a helpful taxonomy of agents within the *ML ecosystem*—that is, a taxonomy of stakeholders who depend on or regularly interact with a computing system developed using Machine Learning. Six kinds of stakeholders are distinguished according to their unique roles within the ML ecosystem (Figure 1), and Tomsett et al. illustrate these stakeholders' roles by invoking the example of a *loan risk assessment system*. Indeed, the risk assessment system is a typical, albeit relatively simple, ML application: A supervised learning algorithm can be used quite straightforwardly to correlate previous applicants' personal data such as income, age, and home address with their eventual ability to repay loans in a timely manner. On the basis of such correlations, the automated loan risk assessment system can take a new individual's personal data as input, and generate an output to estimate the financial risk a bank would incur by accepting that individual's loan application.

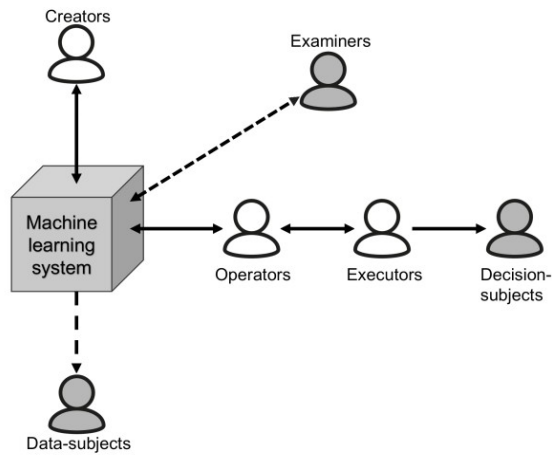

**Figure 1**: The ML ecosystem. **Reproduced from Tomsett et al. (2018).**

In the risk assessor's ecosystem, bank employees are *operators* and/or *executors*. Operators are agents who "provide the system with inputs, and directly receive the system's outputs" (Tomsett et al., 2018, p. 10). These are likely to include bank tellers tasked with entering a particular applicant's personal data into the system, and with receiving the system's output. In contrast, executors are "agents who make decisions that are informed by the machine learning system" (Tomsett et al., 2018, p. 10). These are likely to include back-office employees who rely on the system's output to make data-driven decisions about whether or not to accept a particular application.

Another important agent in the ML ecosystem is the *decision-subject*. In the present example, the decision-subject is the loan applicant: the individual whose data are being processed for the purposes of assessing risk, and who is subject to the executor's final decision. Decision-subjects are distinguished from *data-subjects*, individuals whose personal data are contained in the learning environment. In the loan-application example, data-subjects include previous applicants whose personal data and loan-repayment behavior ground the risk assessor's learned correlations.

*Creators* are the developers of the computing system. These include software engineers responsible for designing the learning algorithm and for selecting the learning environment. They are also likely to include system administrators responsible for maintaining and possibly fine-tuning the system once it has been deployed. Of course, creators might also include hardware engineers tasked with building and maintaining the hardware in which the risk assessor is implemented. That said, many current ML applications are driven by standard-issue hardware components whose workings require no ML-specific knowledge.[6] Perhaps for this reason, Tomsett et al. (2018) do not explicitly consider creators of this kind.

Finally, *examiners* are "agents tasked with compliance/safety-testing, auditing, or forensically investigating a system" (Tomsett et al., 2018, p. 13). Thus, examiners are likely to include regulatory bodies charged with determining that the risk assessment system conforms to, for example, data privacy regulations, anti-discrimination laws, and policies designed to prevent financial fraud. Given legislative innovations such as the EU General Data Protection Regulation (GDPR), examiners are likely to play an increasingly prominent role in future ecosystems.

---

6   Applications that depend on ML-specific hardware components are considered in Section 5.

*3.2. What are the epistemically relevant elements?*

In order to perform their designated roles in the ecosystem, every stakeholder must possess a particular kind of knowledge. More precisely, every stakeholder must possess knowledge of appropriate epistemically relevant elements. Consider again the loan risk assessor. Bank tellers tasked with entering inputs and receiving outputs can only operate the system if they know, for example, something about the INCOME, D.O.B. and HOME_ADDRESS variables (in particular, their type), and that the most important output (the output whose value is to be passed on to the back-office executor) is the value of the RISK variable. In contrast, software engineers charged with developing, maintaining and improving the risk assessor's behavior must be able to identify, characterize, and intervene on the variables—system parameters and/or high-level representations—that mediate the transformation of inputs to outputs. Perhaps surprisingly, most other agents are less likely to be concerned with the system's variables than with the environmental features that are represented by those variables. In particular, although executors, examiners, data- and decision-subjects may not need to know much about variables such as INCOME, D.O.B., HOME_ADDRESS, and RISK, but will have to know something about (their own or their customers') income levels and demographic information, and eventually, about the level of financial risk associated therewith.

Given that they must possess a particular kind of knowledge, opacity—a *lack* of knowledge—prevents every kind of stakeholder from performing their designated role in the ecosystem.[7] If a bank teller does not know that an applicant's date of birth must be entered YYYYMMDD rather than DDMMYYYY, he or she will be unable to correctly operate the system so as to generate meaningful outputs. If a back-office executor does not know that an output value of 0.794 designates a relatively high level of financial risk, he or she will be unable to make an appropriate decision with respect to a particular loan application. If current applicants (i.e., decision-subjects) are unable to acquire knowledge of the factors that contribute to particular decisions, they might not be able to exercise their GDPR *right to explanation* (Goodman & Flaxman, 2016; but cf. Wachter, Mittelstadt, & Floridi, 2017). Similarly, if previous applicants (i.e., data-subjects) are unable to acquire knowledge of the personal data that is stored in the learning environment, they will not be in a position to exercise their GDPR *right to information*. If examiners such as lawyers or regulatory bodies cannot discern whether the loan risk assessment system has learned to correlate a foreign place of birth with high level of financial risk, they will be unable to identify, and if necessary sanction, possible discrimination. Finally, software engineers who do not know the variables that mediate causally between a system's inputs and outputs—be they individual system parameters or high-level representations—will be unable to efficiently modify the system's performance so as to bring it in line with the lawyer's legal recommendations.

---

[7] Some commentators deny that opacity prevents stakeholders other than creators from fulfilling their roles (see, e.g., Zerilli, Knott, Maclaurin, & Gavaghan, 2018). However, these commentators fail to recognize the diversity of stakeholders within the ML ecosystem, and thus, the different senses in which computing systems may be opaque, as well as the different reasons for rendering these systems transparent.

Where opacity is the problem, transparency is the solution. Thus, stakeholders should seek explanations—they should render the relevant computing system transparent. Of course, given that the knowledge required to perform their roles differs between stakeholders, so too will the knowledge that should be sought. In particular, different stakeholders should seek knowledge of different kinds of epistemically relevant elements. But although the appropriate EREs may be apparent enough in the context of the loan-application example, in order to extend the discussion to other (potentially more complex) examples—and thus, to understand in a more general sense of what is required to render computing systems transparent—it will be necessary to attain a deeper understanding of the different kinds of knowledge that different stakeholders require in order to fulfill their designated roles within the ML ecosystem.[8]

*3.3. Toward a Marrian framework for Explainable AI*

One way of attaining such a deeper understanding is to adopt a more general explanatoryframework. Consider David Marr's (1982) *levels of analysis* account, which specifies norms that should be satisfied in order to explain the behavior of cognitive systems as diverse as the human visual system and the sensorimotor system of the house fly. Notably, despite having been developed almost four decades ago, Marr's account remains influential in cognitive science today, and there are reasons to believe that it can serve double-duty as a normative framework for Explainable AI in the future. For one, like many biological cognizers, the computing systems being developed in Artificial Intelligence can be viewed as information-processing systems in which inputs are systematically transformed into outputs.[9] For another, like the computers being programmed using Machine Learning, biological cognizers are opaque in the sense that we still do not know exactly why they do what they do or how they work (Zerilli et al., 2018).

---

8  Although the present discussion distinguishes between the explanatory requirements of different kinds of stakeholders, there may also be differences between the explanatory requirements of any two individuals, even if they are stakeholders of the same kind. In particular, differences in the background knowledge, training, and preferences of individuals are likely to influence the receptivity of these individuals to specific kinds of explanations. But although such individual differences are undeniably interesting and relevant, the present discussion will not consider them further, and will instead only focus on the the differences that arise between different kinds of stakeholders due to their distinct roles.

9  Notably, in order to illustrate and motivate his framework, Marr himself did not only invoke biological cognizers, but also invoked engineered computing systems such as the supermarket cash register. Indeed, the agents who regularly interact with the cash register are closely analogous to the agents in the ML ecosystem: the supermarket employees who operate the register are like operators and/or executors; customers are like decision subjects; the software or hardware engineers charged with constructing and maintaining the cash register are like creators; financial regulatory bodies such as tax auditors are examiners. Although there exist important disanalogies (e.g. the supermarket cash register has no data-subjects), the fact that Marr's framework is in part motivated by its applicability to engineered systems can be viewed as an additional reason to believe that may also be useful when articulating a normative framework for Explainable AI.

That said, the most important reason for thinking that Marr's account of explanation in cognitive science can serve double-duty as a normative framework for Explainable AI is that the account is sufficiently multi-faceted to illuminate the subtle differences between the knowledge required by different stakeholders. Indeed, Marr's account centers on the answering of several different *questions* at three distinct *levels of analysis* (McClamrock, 1991; Shagrir, 2010; Zednik, 2017). Insofar as different stakeholders can be thought to ask different kinds of questions, the characteristic answers described by Marr can be used to better understand the different kinds of EREs these stakeholders should invoke in order to render a computing system transparent. Indeed, by adopting Marr's account to evaluate the explanatory contributions of analytic techniques from Explainable AI, it will be possible to determine exactly which questions these techniques are capable of answering, and thus, for which stakeholders and to what extent the analyzed systems can eventually be rendered transparent.

## 4. Description and Interpretation: The Computational Level

*4.1. Questions about* what *and* why

Marr's *computational level of analysis* centers on questions about *what* a system is doing, and on questions about *why* it does what it does. These questions are intimately related, but importantly different. Questions about *what* a particular system is doing call for a description of that system's overall behavior in terms of, for example, a transition function $f$ in which 'input' states are mapped onto corresponding 'output' states (Figure 2). In the context of the loan risk assessment system, what-questions are answered by specifying the value of the RISK variable that is generated for any particular combination of input variables such as INCOME, D.O.B., and HOME_ADDRESS.

In contrast, questions about *why* a system does what it does are questions about that behavior's "appropriateness" within some particular environment (Marr, 1982, p. 24f). That is, these questions call for an interpretation of the system's behavior in terms of recognizable features of the environment—be this the learning environment that (together with the learning algorithm) determines the system's behavior, or the behavioral environment in which the system is eventually deployed. Insofar as what-questions are answered by specifying a transition function $f$ that maps 'input' states onto corresponding 'output' states, why-questions concern the environmental regularity or correlation $f'$ that obtains between the inputs and outputs themselves. Specifically, answering a why-question involves showing that there exists a correspondence between $f$ and $f'$ (Shagrir, 2010. See also Figure 2). Thus for example, in the loan-application scenario, why-questions are not answered by describing the values of variables such as INCOME, D.O.B., HOME_ADDRESS, and RISK, but rather by interpreting these values in terms of features such as a particular applicant's income, date of birth, home address, and level of financial risk.

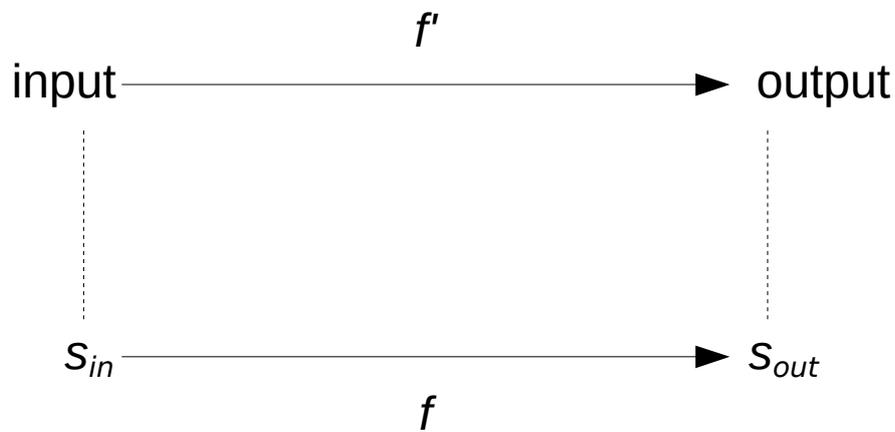

**Figure 2.** $S_{in}$ and $S_{out}$ are the 'input' and 'output' states of the computing system, respectively; input and output are the corresponding features of the environment. The solid arrow at the bottom designates the overall behavior of the system, presumably realized in some causal process. The dotted arrow designates a representation (or other kind of correspondence) relationship between the system and its environment. What-questions concern the transition function $f$ from $s_{in}$ to $s_{out}$. Why-questions, in contrast, concern the relationship $f'$ between input and output. **Adapted from Shagrir (2010).**

It can be helpful to view the correspondence between systems and their environments semantically, that is, in representational terms. The system's 'input' and 'output' states may be thought to represent the input received from the environment and the output that is generated in response. Likewise, the state-transition $f$ might be thought to track the regularity or correlation $f'$ that obtains between those inputs and outputs. Although more can and should be said about the correspondence that obtains between a computing system and its environment (for discussion see, e.g., Shagrir, 2010), for current purposes it suffices to say that whereas what-questions concern local properties of the computing system itself—often, its representational vehicles—why-questions concern features of the surrounding environment—the corresponding representational contents.

*4.2. Operators, executors, examiners, data- and decision-subjects*

This brief presentation of the computational level can already be used to align different stakeholders in the ML ecosystem with distinct questions within Marr's account. For this reason, it can be used to better understand the kinds of EREs these agents should identify in order to render computing systems transparent.

Insofar as they are tasked with entering inputs and receiving outputs, operators are most likely to seek answers to questions about *what* a computing system is doing. That is, their explanatory requirements are satisfied—the system is rendered suitably transparent—by describing the inputs that must be entered and the outputs that are generated. In contrast, executors and examiners are far more likely to be concerned with why-questions in which a system's behavior must be interpreted. To wit, although a back-office executor in the loan-application scenario must know that the risk assessment system computes a value of 0.794, his or her most important task is to interpret that value as an indicator of significant financial risk. Similarly, it is an examiner's duty to determine whether a particular assessment has been generated legitimately, or because the system discriminates by associating a foreign place of birth with a high level of financial risk.

Data- and decision-subjects are similarly concerned with questions about *why* rather than with questions about *what*. Since coming into force in May 2018, the GDPR empowers data- and decision-subjects to seek answers to why-questions, but does not similarly extend to what-questions. In particular, the GDPR *right to information* allows data-subjects to know which personal information is represented, but not how that information is entered, represented, and manipulated. Similarly, the GDPR *right to explanation* may allow decision-subjects to know which factors contributed to a particular outcome (e.g., a low income may lead to a denied application), but not to the particular way in which those factors are represented, or how the outcomes are actually calculated.[10] Notably, the rationale for focusing on questions about *why* rather than on questions about *what* (or on questions about *how*, see Section 5) is compelling: whereas my personal data belong to me, the data-structures and processes that are actually used to represent and manipulate those data are the property of the AI service provider.

In summary, the distinction between what-questions and why-questions at the computational level captures an important distinction between two different ways of rendering a computing system transparent, each of which is appropriate for different stakeholders within the ML ecosystem. Whereas operators typically seek to render a system transparent by asking *what* it does and describing its 'input' and 'output' states, many other agents do so by asking *why* it does what it does and interpreting those states in terms of environmental features and regularities. Perhaps surprisingly, given the metaphorical underpinnings of "opacity" and "black boxes", this means that for several stakeholders in the ML ecosystem, rendering a computing system transparent does not involve "looking inside" the system at all, but rather "looking out", at the environment in which that system's behavior is learned and performed.

*4.3. Input heatmapping*

Several XAI techniques can be used to answer questions about *what* and *why*. One such technique is *input heatmapping*, which highlights the features of a system's input that are particularly relevant to, or predictive of, its output. Some of the most compelling examples of this technique come from the domain of machine vision, in which deep neural networks are used to classify pixelated input images according to the people, objects, properties, or situations they depict.[11] There, input heatmapping typically involves the generation of visualizations—i.e., heatmaps—that emphasize the particular pixels or pixel regions that are most responsible for a particular classification (Figure 3).

---

10   It is a point of contention to what extent the GDPR right to explanation constitutes a right at all, and if so, what it guarantees (Wachter, Mittelstadt, & Floridi, 2017). Although the former is a legal question that goes beyond the scope of the present discussion, the latter is a normative question that may benefit from the present discussion. In particular, Goodman & Flaxman (2016) argue that the GDPR grants data-subjects (and decision-subjects) the right to acquire "meaningful information about the logic involved." However, what exactly is meant by "logic" in this context remains unclear. The present discussion implies that the relevant stakeholders should be primarily concerned with why-questions, and that in this sense, the "logic" should be specified in terms of learned regularities between environmental factors. Moreover, the present discussion suggests that AI service providers may, for legal reasons, be compelled to deploy XAI techniques capable of answering why-questions.

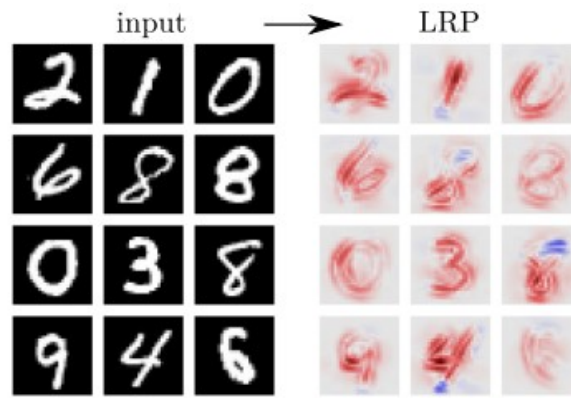

**Figure 3.** Input heatmaps (right) for various input images (left), generated by applying the method of layer-wise relevance propagation (LRP) to a deep neural network capable of recognizing handwritten digits. **Reproduced from Montavon et al. (2018)**.

One concrete approach for developing heatmaps for artificial neural networks is *Layer-wise Relevance Propagation* (LRP, Montavon et al., 2018). This method deploys a subroutine of the popular backpropagation learning algorithm, in which individual unit activations and connection weights are used to calculate the responsibility that the individual units of an "upstream" layer $l_i$ bear for producing particular levels of activity in the subsequent "downstream" layer $l_{i+1}$. Given a particular classification at the network's output layer, this subroutine can be deployed in a layer-wise fashion until responsibility-values are calculated for every unit in the network's input layer. Insofar as these units correspond to, for example, pixels of a particular input image, these responsibility-values can be used to generate a heatmap that highlights the pixels or pixel regions that bear the greatest responsibility for the final classification.

Input heatmaps provide particularly fine-grained answers to questions about *what* a computing system is doing. Recall that questions of this sort are answered by specifying the transition function *f* that obtains between a system's 'input' and 'output' states. Input heatmaps can help specify *f*, not in terms of the 'input' state as a whole (e.g., a whole pixelated image), but in terms of a limited number of elements within that state (e.g., individual pixels). Such fine-grained answers to what-questions are particularly useful for operators. In particular, they greatly enhance an operator's ability to identify the most likely sources of error. If an error is detected in the 'output' state (e.g., the RISK value is inappropriately high), it is more likely to result from high-responsibility elements of the 'input' state (e.g., a wrongly-formatted D.O.B. value) than from low-responsibility elements (e.g., a misspelled surname). That said, such fine-grained answers to what-questions might also be exploited by malicious operators such as hackers. Indeed, input heatmaps can be used to design *adversarial inputs* that appear typical to human observers, but that nevertheless produce radically divergent (and thus, potentially exploitable) outputs due to minor changes to some high-responsibility elements (Szegedy et al., 2013).

---

11  Although machine vision may be the domain in which input heatmaps are most intuitive, they may also be used in other domains. For example, input heatmaps may be constructed for audio inputs, highlighting the moments within an audio recording that are most responsible for classifying that recording by musical genre. Moreover, although LRP is specifically designed to work with artificial neural networks, other methods can be used to generate input heatmaps for other kinds of systems. Thus, input heatmapping can be viewed as a general-purpose XAI technique for answering what- and why-questions.

Notably, input heatmaps can also be used by creators. Creators are regularly tasked with improving or otherwise changing a system's overall behavior, and thus, with changing the system's transition function *f*. An input heatmap's fine-grained answers to what-questions can be used by creators to identify the necessary changes. Thus for example, in order to maximize a system's processing efficiency and minimize its memory load, an input heatmap might be used to determine whether certain aspects of the input (e.g., pixels at the edge of an image) can be ignored, or whether the memory required to store individual inputs can be reduced by, for example, zeroing the values of low-responsibility elements. Similarly, creators seeking to minimize a system's susceptibility to adversarial inputs could deploy input heatmaps to, for example, determine that the responsibility for final classifications should be distributed more evenly across all 'input' elements. Of course, although creators may invoke input heatmaps to determine the particular changes that must be made, they will normally have to use other techniques—in particular, techniques capable of answering how-questions—in order to know how those changes can actually be achieved.

While input heatmapping can be used by stakeholders seeking to better understand *what* a system is doing, this technique can also be used by stakeholders interested in knowing *why* the system does what it does. Recall that why-questions are answered by specifying a regularity or correlation *f'* that obtains between features of the environment, and by showing that this correlation is tracked by the system's transition function *f*. Input heatmaps can be used to answer why-questions if the environmental features that participate in *f'* can be discerned by inspecting the map—that is, if the highlighted elements of the 'input' state together "look like" some recognizable feature of the environment. Consider again the input heatmaps in Figure 3. The highlighted pixels visually resemble handwritten digits. In particular, the highlighted pixels in the upper-left heatmap visually resemble a handwritten 2 rather than, for example, a handwritten 7. This fact answers a question about *why* the upper-left input image outputs '2' rather than '7': the output is '2' *because* the image depicts a 2. Stated more generally, the input heatmaps in Figure 3 show that the system does what it is in fact supposed to do, namely, detect and classify handwritten digits.

In order to better appreciate the importance of answering why-questions in this way, it is worth contrasting the above example with one in which the computing system does not in fact do what it is supposed to do. Consider a well-known historical (albeit probably apocryphal) example in which a neural network learns to visually distinguish enemy tanks from friendly ones.[12] Although the network quickly learns to categorize images of tanks, it does so by tracking an accidental correlation between tank allegiances and weather patterns: whereas the images of friendly tanks were all taken on a sunny day, the enemy tanks were photographed under cloud cover. For this reason, although the system correctly classifies images (*what*), its reasons for doing so (*why*) have nothing at all to do with tanks!

In this example, heatmapping techniques such as LRP would be likely to produce visualizations in which the highlighted pixels together resemble clouds in the background, rather than tanks in the foreground. Several different stakeholders would be likely to benefit from such visualizations: executors at military HQ who must ultimately decide whether or not to shoot at a particular tank, but also examiners at the International Criminal Court tasked with determining whether the resultant action ought to be considered a war crime. Moreover, tank crews who make up the decision-subjects of the tank-classification system could rest easy in their knowledge that the system can be trusted, and that they are for this reason less likely to perish in a data-driven barrage of friendly fire.

Before moving on to other kinds of questions, it is important to mention a significant limitation of the heatmapping technique. Input heatmaps can be used to answer why-questions when the highlighted elements together "look like" some recognizable feature of the environment. As was already suggested in Section 2, however, Machine Learning methods are renowned for their ability to identify and track subtle as well as complex features and correlations in the learning environment, many of which may not be easily recognized or labeled by human observers. In such cases, the utility of input heatmapping is likely to be limited, and other XAI techniques may have to be invoked. One such technique might be *Local Interpretable Model-agnostic Explanations* (LIME, Ribeiro et al., 2016), which can be used to simplify the transition function $f$ (which is often nonlinear and therefore difficult to interpret) with a linear approximation that is more readily interpretable by human observers. Similarly, *Local Rule-Based Explanations* (LORE, Guidotti et al., 2018) are designed to approximate limited domains of $f$ by sets of comprehensible decision-rules. Although these different approximations always only capture a system's behavior for a limited range of inputs, and although as approximations they always bear the risk over-simplification and misrepresentation, they may nevertheless prove useful for answering what- and why- questions even when input heatmapping fails.

---

[12] Gwern Branwen maintains a helpful online resource on this particular example, listing different versions and assessing their veracity: https://www.gwern.net/Tanks (retrieved January 25[th], 2019).

# 5. Intervention: The Algorithmic and Implementational Levels

*5.1 Questions about* how *and* where

In Marr's account of explanation in cognitive science, the levels "below" the computational level of analysis are the *algorithmic* and *implementational levels*. The algorithmic level centers on questions about *how* a system does what it does. Insofar as the system's behavior is described using a transition function *f* from an 'input' state to an 'output' state, the algorithmic level aims to uncover the mediating states $s_1, s_2,...s_n$ and state-transitions $s_i \rightarrow s_j$ that appropriately connect 'input' and 'output' (Figure 4). Put differently, the algorithmic level is concerned with uncovering the *program* that executes the overall transition *f*, and to thereby compute or approximate *f'* (Shagrir, 2010).[13]

In contrast, the *implementational level of analysis* centers on questions about *where* the program described at the algorithmic level is realized. Where-questions concern the physical components $p_1, p_2,..p_m$ in which states are realized and state-transitions are performed (Figure 4). Thus, the implementational level is concerned with the hardware components that are involved in executing the program for *f*.

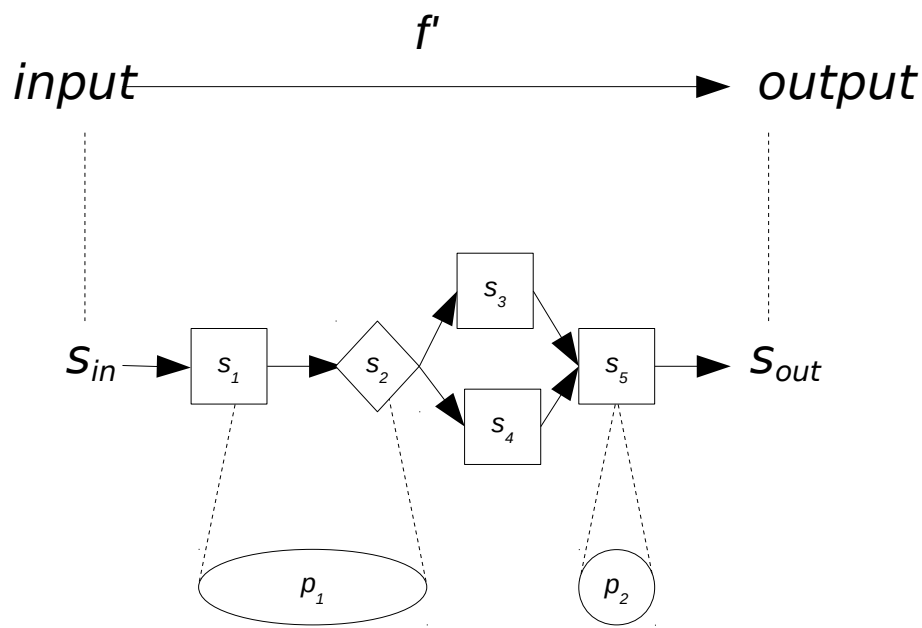

**Figure 4.** As in Figure 3, $S_{in}$ and $S_{out}$ are the 'input' and 'output' states of the computing system, respectively; input and output are the corresponding features of the environment, and the dotted arrow designates a representation (or other kind of correspondence) relationship between the system and its environment. The squared shapes and arrows in the middle designate the individual states and state-transitions that are performed to compute or approximate *f*, the description of which answers how-questions. The circles at the bottom indicate the physical realizers of individual states and/or state-transitions, the identification of which answers where-questions.

---

13  The program that mediates between 'input' and 'output'—the program being executed—is not to be confused with the learning algorithm that is used to develop (i.e., *to* program) the system in the first place.

Whereas what- and why-questions can both be answered by "looking out" at a system's surrounding environment, how- and where-questions can only be answered by "looking inside", at the functional or physical variables, structures, and processes that intervene between the system's inputs and outputs. For this reason, these questions are particularly important for the purposes of intervening on a system's behavior. Knowledge of a state $s_i$ or a transition $s_i \rightarrow s_j$ that mediates the overall transition between 'input' and 'output' can be used to influence that overall transition by changing either one of $s_i$ or $s_i \rightarrow s_j$. Likewise, knowledge of the fact that either $s_i$ or $s_i \rightarrow s_j$ is physically realized in some physical structure $p_k$ can be used to achieve the same goal by, for example, replacing $p_k$ with some other physical structure $p_l$, or by removing $p_k$ altogether.[14]

In order to better understand the way in which creators should go about answering how- and where-questions in the Machine Learning context, it is instructive to first consider the way these questions are answered in cognitive science. There, answers to how-questions are typically delivered by developing *cognitive models* which describe the processes that govern a particular system's behavior. Notably, the relevant processes are only rarely described in terms of "brute causal" interactions between neuronal structures. More commonly, they are described in terms of the numeric calculation of values, or the step-wise transition between states (Busemeyer & Diederich, 2010). The most important advantage of such descriptions is that they are able to capture a system's abstract mathematical properties. However, in some cases they also afford relatively straightforward semantic interpretations (Fodor, 1987). That is, certain variables or states of a cognitive model might be said to represent specific features of the environment, so that changes in those variables or transitions between those states capture changes in the things being represented. That said, representational interpretations may not always be forthcoming—for example, because it is unclear which features of the environment are actually being represented (Ramsey, 1997)—nor useful—in particular, when the non-representational description is deemed sufficiently useful for the purposes of interpreting, predicting, and intervening on the system's behavior (Chemero, 2000).

In turn, where-questions in cognitive science are typically answered by *localizing* the elements of a cognitive model—i.e., its individual states, state-transitions, variables, or calculations—in specific physical structures such as neurons, neural populations, or brain regions (Piccinini & Craver, 2011; Zednik, 2017). Although cognitive scientists had long denigrated the explanatory relevance of where-questions (Pylyshyn, 1984), it is now widely acknowledged that a proper understanding of the brain is critically important for the purposes of explaining the mind (Shallice & Cooper, 2011). That said, it would be a mistake to think that localization in cognitive science is typically "direct" in the sense of affording simple one-to-one mappings between the individual steps of a process and the parts of an underlying physical structure (Bechtel & Richardson, 1993). Indeed, the functional boundaries between the elements of cognitive models frequently cut across the physical boundaries between physical structures in the brain (Stinson, 2016). Although this should not be taken to indicate that the answering of where-questions is impossible or unimportant, it does show that close attention should be paid to the sense in which, for example, the states and state-transitions of a particular process—and thus, possibly, the corresponding representations and representation-manipulations—might be *distributed* (Clark, 1993).

---

14 Although there is a clear sense in which interventions can also be achieved by modifying a system's inputs—a different $s_{in}$ will typically lead to a different $s_{out}$—interventions on the mediating states, transitions, or realizers are likely to be far more wide-ranging and systematic.

*5.2. Creators*

How might this brief foray into the explanatory norms and practices of cognitive science be relevant to Machine Learning? Like scientists working to explain the behavior of biological cognizers, the creators of an ML-programmed computing system are frequently preoccupied with questions about *how* and *where*. On the one hand, software engineers and system administrators tasked with developing, maintaining, fixing, and generally improving a system's behavior need to know not only *what* that system does, but also *how* it does it. On the other hand, hardware engineers charged with building and maintaining the hardware in which the system is implemented must know *where* certain processes could be, or are actually, localized. Notably, much about the way in which how- and where-questions are answered in cognitive science can be used to better understand the way these questions can and should be answered in Explainable AI.

Consider how-questions first. Just as in cognitive science it is only rarely useful to look for "brute-causal" interactions between neuronal structures, in the Machine Learning context it is only rarely useful to cite the values of individual learnable parameters such as a neural network's unit activations or connection weights. Indeed, as has already been discussed in Section 2, their high-dimensional complexity makes many ML-programmed computing systems unpredictable even with complete knowledge of the underlying parameter values. Moreover, there is often no way of knowing in advance whether an intervention on a single parameter will change the relevant system's behavior entirely, or else affect it in a way that is mostly or entirely imperceptible.

Like in cognitive science, therefore, suitable answers to how-questions in Explainable AI will in most cases describe abstract mathematical properties of the system whose behavior is being explained. To better understand the kinds of properties that might be sought, consider once again the loan risk assessment system from above. In order to track statistical correlations between previous applicants' personal data and their ability to repay loans, the risk assessment system is unlikely to categorize new applicants on the basis of simple linear combinations of inputs such as age and income. Rather, the system is more likely to deploy a taxonomy of abstract categories in which the inputs are combined nonlinearly (see also Buckner, 2018). For illustrative purposes, it can help to assume that these categories approximately correspond to folk-psychological character traits such as *honest, unscrupulous, high self-control, persevering,* or *foolhardy*. Although an applicant who once was the victim of a ponzi scheme may appear inconspicuous to a bank employee, the automated risk assessment system might, through a nonlinear combination of data points, nevertheless classify that applicant as being foolhardy, and for this reason, risky. In order to properly understand *how* this system does what it does—and thus, in order to potentially intervene on the system's behavior—creators would almost certainly profit from identifying the system's learned categories, and from characterizing the role these categories play in the generation of particular outputs.

Now, consider where-questions. Recall that questions of this kind were long denigrated in cognitive science. Because many ML applications are driven by standard-issue hardware components, it may be tempting to dismiss where-questions as being similarly irrelevant to Explainable AI. But it is important to resist this temptation. Indeed, a growing number of ML applications are driven by ML-specific hardware components. For example, the artificial neural networks used for visual processing in self-driving cars are increasingly implemented on neuromorphic hardware devices that have considerable advantages with respect to speed and power consumption (Pfeiffer & Pfeil, 2018). In this sense, where-questions are hugely important at least for creators tasked with choosing the hardware in which to implement a particular computing system. But where-questions can also be important for creators tasked with maintaining, repairing, or improving a computing system once it has been built and deployed. Knowing the physical location in which certain kinds of data are stored and processed can allow creators to selectively replace hardware components so as to improve the speed or efficiency of the system as a whole. Moreover, in certain scenarios, it may even be important to know where data are processed *after* the system has stopped working. For example, in the aftermath of a fatal accident involving a self-driving car, it may be necessary to extract the data that was processed in the moments immediately preceding the accident, even if the system is no longer operational.[15] In such scenarios, a creator's ability to answer questions about *where* certain operations were performed may be instrumental for answering further questions about *what* the system was doing, as well as *why* and *how*.

Surprisingly perhaps, where-questions may sometimes even be important to stakeholders other than creators. Although many computing systems are colloquially said to be realized "in the cloud", what is actually meant is that they are implemented in a device (or a cluster of devices) that is physically far-removed from the stakeholders in the relevant ecosystem. Given the differences in the ways different countries regulate the storage of information, data- and decision-subjects alike may want to know the particular jurisdiction under which—i.e., *where* in the world—their data is being stored and processed. In a related way, examiners may be tasked with developing and enforcing legislation that limits the extent to which data can be exported. Thus, whereas where-questions are typically within the purview of creators, there are situations in which these questions may also become important to other stakeholders within the ML ecosystem.

*5.3. Feature-detector identification*

One technique for answering how- and possibly even where-questions involves the identification of *feature-detectors*. Just as what- and why-questions can be answered by highlighting features of the input that bear a high responsibility for the production of certain outputs, how-questions can sometimes be answered by highlighting those features of the input that are most responsible for activity in certain mediating variables. Insofar as the relevant variables are *sensitive* to a particular feature (i.e., respond reliably when the feature is present), relatively *unique* (i.e., no other mediating variables are similarly sensitive), and *causally efficacious* (i.e., they significantly influence the system's overall output), those variables can be viewed as *feature-detectors*. Identifying a system's feature-detectors—assuming there are any—and exploring their influence on the system's overall behavior serves well for the purposes of answering questions about *how* that system works, and sometimes, even about *where* the relevant feature-detecting operations are carried out.

---

15  Curiously, in such scenarios a computing system's hardware components become analogous to the "black box" voice-recorders used on commercial airliners.

Consider a recent study due to Bau et al (2018). This study considers *generative adversarial networks* (GANs) that are capable of producing photorealistic images of scenes depicting, for example, christian churches (Figure 5a). The aim of the study is to identify feature-detectors and explore the systematic effects of surgical interventions on these detectors.[16] To this end, Bau et al. "dissect" the relevant networks to identify the units or unit clusters that are sensitive and unique with respect to recognizable features such as trees (Figure 5b). Subsequently, they determine the causal efficacy of these (clusters of) units by performing a series of interventions such as activating or ablating (i.e., de-activating) the relevant units. Indeed, through these kinds of interventions the authors are able to systematically control the presence or absence of features in the generated images, without compromising the overall image quality (Figures 5c and 5d).

---

16  Strictly speaking, because the aim of the GANs in this study is not detection but *generation*, the relevant units might more appropriately be called feature-*generators*.

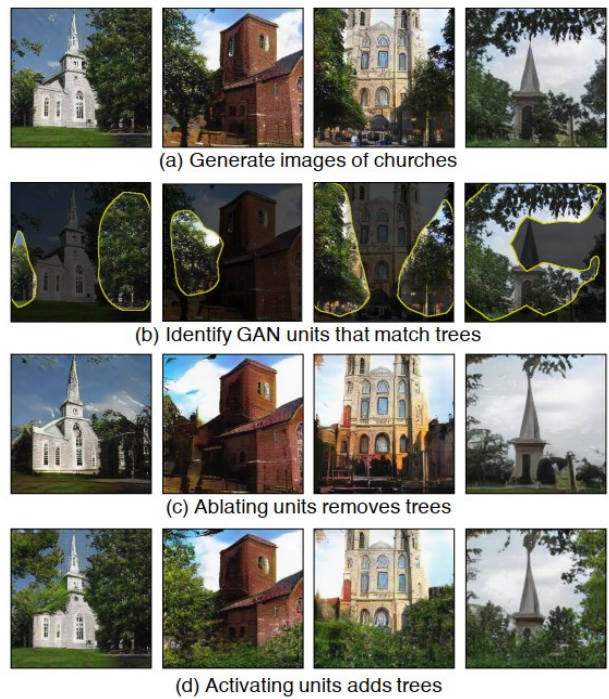

**Figure 5.** Images generated by a generative adversarial network: (a) generated images of christian churches and their surroundings; (b) features detected by the feature-detector for trees; (c) images generated after ablating the feature-detector for trees; (d) images generated after activating the feature-detector for trees. **Reproduced from Bau et al. 2018**.

Bau et al.'s study exhibits many of the hallmarks of well-answered how-questions. For one, it shows that interventions on feature-detectors can be used to repair or otherwise improve the relevant system's performance—a typical task for creators. For example, by identifying and subsequently ablating the feature-detectors for unwanted visual artifacts such as textured patterns where there should be none, they are able to successfully remove those artifacts from the generated images and therefore improve the system's overall performance. For another, most of the features being detected by the networks in the study are robust with respect to nuisance variations in color, texture and spatial orientation. Thus, the GANs appear to learn just the kinds of high-level representations that one would expect to find in high-dimensional complex systems programmed using Machine Learning. Indeed, the authors even advance the hypothesis that the networks' representations resemble the conceptual representations that are used by human brains (see also: Buckner, 2018).

As this example shows, the identification of feature-detectors is an effective technique for answering how-questions in the ML ecosystem. That said, it is worth considering the extent to which this technique might also be used to answer where-questions. Indeed, insofar as a network's feature-detectors are concentrated in a relatively small number of units and those units are implemented in neuromorphic hardware components, the identification of feature detectors will answer questions

Page 21

about *how* and *where* simultaneously. Indeed, in such cases, modifications to the networks' overall behavior could equally be achieved by, for example, removing or replacing the hardware components that implement specific detectors.

*5.4. Diagnostic classification*

Feature-detectors can be invoked to answer questions about *how* and even *where,* but only when the responsibility for generating particular outputs is concentrated in a relatively small number of system variables (e.g., a small number of network units). In contrast, they cannot normally be invoked when the responsibility is distributed accross a large number of variables (e.g., a layer or network as a whole). Indeed, decades-long discussions of connectionist modeling methods in cognitive science suggest that neural networks and similarly high-dimensional systems are very likely to exhibit this kind of distributed responsibility (Smolensky, 1988), and thus, are likely to deploy distributed representations (Clark, 1993). For this reason, investigators in Explainable AI have good reason to develop alternative techniques that can be used to answer how-questions even when no clearly circumscribed feature-detectors can be found.

Consider a recent study from computational linguistics, in which Hupkes et al. (2018) explore the capacity of different networks to evaluate expressions with nested grammatical structures. It has long been known that *simple recurrent networks* (SRNs, Elman, 1990), like other networks with recurrent feedback connections, perform well when confronted with tasks of this nature. What remains unclear, however, is exactly *how* these networks do what they do, and in particular, how they store and deploy information over extended periods of time. To wit, assuming that a nested arithmetic expression such as "(5-((2-3)+7))" is processed from left to right, some symbols encountered early (e.g., the '5' and the leading '-') will have to be evaluated only after other symbols are encountered later.

Hupkes et al.'s challenge is to determine whether networks capable of evaluating such nested expressions do so by following either one of two strategies: a *recursive* strategy in which parenthetical clauses are evaluated only once they have been closed (2-3=-1; -1+7=6; 5-6=-1), or a *cumulative* strategy, in which parentheses are removed from left to right while appropriately "flipping" the operators between '+' and '-' (5-2=3; 3+3=6; 6-7=-1). In order to answer this question about *how* the relevant networks do what they do, Hupkes et al. deploy *diagnostic classifiers*: secondary networks that take as inputs the primary network's hidden-unit activations while a particular symbol is being processed, and that generate as output a value that can be compared to a prior hypothesis about the information that *should* be represented at that moment. In the present example, the diagnostic classifiers' outputs are compared to the information that should be represented if the recurrent network were to follow either one of the relevant strategies. For example, after processing the '3' in the arithmetic expression "(5-((2-3)+7))", a recurrent network that adheres to the recursive strategy should represent an intermediate sum of -1, whereas one that follows the cumulative strategy should represent a 6. Indeed, Hupkes et al. find that the diagnostic classifiers generate outputs more in line with the cumulative strategy than with the recursive strategy

(Figure 6).

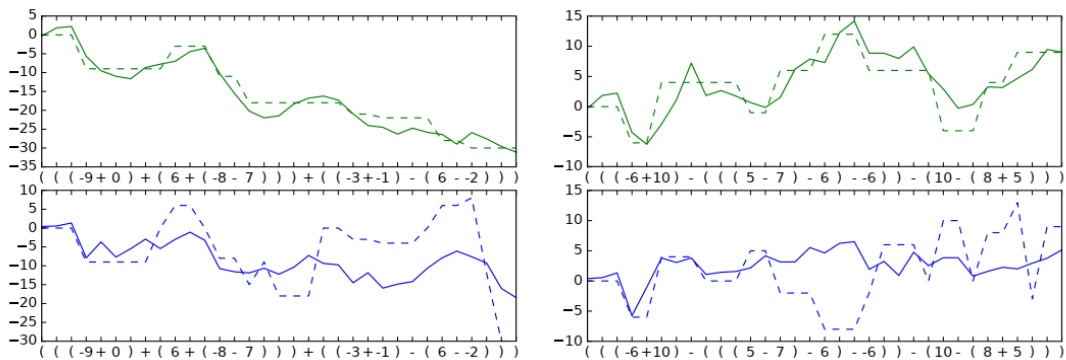

**Figure 6.** Diagnostic classifier outputs (y-axis) for the cumulative (top) and recursive (bottom) strategies over the course of two distinct expressions (x-axis). The classifiers' outputs (solid line) are compared to the prior hypotheses (dashed line), revealing an overall closer fit to the cumulative than to the recursive strategy. **Reproduced from Hupkes et al. (2018).**

Like feature-detectors, diagnostic classifiers can be used to answer questions about *how* a particular system does what it does. More precisely, they can be used to determine which information is represented by a system when it receives a particular input, and thus, how that network processes information as the inputs change over time. In a sense, therefore, diagnostic classifiers can be used to trace the particular computer program—in this case understood as a series of transitions between information-bearing states—that is executed by networks that are capable of solving complex problems in AI.

Diagnostic classification has at least one important advantage over techniques that center on feature-detectors, but also some significant disadvantages. On the one hand, diagnostic classifiers do not require that the representations in the relevant system be contained in a small number of variables. For this reason, they appear well-suited for answering how-questions even when—as is often likely to be the case—networks solve AI problems by manipulating distributed representations. On the other hand, diagnostic classifiers may be thought to be of comparatively limited explanatory value with respect to how-questions insofar as they do not afford the ability to intervene on these systems' behavior. Whereas feature-detectors can be ablated or activated, and the resultant effects can be recorded, it is not clear how a diagnostic classifier's outputs can be used to systematically modify a system's behavior. Moreover, whereas in certain cases feature-detectors may be cited to answer where-questions in addition to answering how-questions, diagnostic classifiers are unlikely to address the former other than by indicating that certain representations are realized somewhere within the system as a whole.

## 6. Conclusion

This discussion has sought to develop a normative framework for Explainable Artificial Intelligence—a set of requirements that should be satisfied for the purposes of rendering opaque computing systems transparent. Beginning with an analysis of 'opacity' from philosophy of science, it was shown that opacity is agent-relative, that that rendering transparent involves acquiring knowledge of epistemically relevant elements. Notably, given their distinct roles in the Machine Learning ecosystem, different stakeholders were shown to require knowledge of different (albeit equally "real") epistemically relevant elements. In this sense, although the opacity of ML-programmed computing systems is traditionally said to give rise to the Black Box Problem, it may in fact be more appropriate to speak of *many* Black Box Problems. Depending on who you are and how it is that you interact with an ML-programmed computer, that computer will be opaque for different reasons, and must be rendered transparent in different ways.

Explainable Artificial Intelligence is in the business of developing analytic techniques with which to render opaque computing systems transparent, and thus, to allow the stakeholders in an ML ecosystem to better develop, operate, trust, examine, and otherwise interact with ML-programmed computing systems. But although many powerful analytic techniques have already been developed, not enough is known about when and how these techniques actually explain—that is, when and in which sense these techniques render computing systems transparent. Here, the present discussion sought inspiration in cognitive science. As one of the most influential normative frameworks for evaluating the explanatory successes of analytic techniques in cognitive science and neuroscience, Marr's *levels of analysis* account was used to determine what it takes to satisfy the distinct explanatory requirements of different stakeholders in the ML ecosystem. By aligning these stakeholders with different kinds of questions—specifically, questions about *what*, *why*, *how*, and *where*—and by specifying the kinds of EREs that should be invoked in order to answer these questions, it was possible to develop a normative framework with which to evaluate the explanatory contributions of analytic techniques from Explainable AI.

Finally, by reviewing a series of illustrative examples, it was argued that many opaque computing systems can already be rendered transparent in the sense required by specific stakeholders. This review demonstrated that it is increasingly possible to not only answer operators' questions about *what* a computing system is doing, but to also answer data- and decision-subjects' questions about *why*, as well as creators' and examiners' questions about *how* the system does what it does, and in certain circumstances, even questions about *where*. Thus, Explainable AI appears to be well-equipped for answering the questions that are most likely to be asked by different stakeholders, and thus, for finding solutions to the many Black Box Problems.

That said, it is important to recognize that the analytic techniques that have thus far been developed have certain characteristic limitations—as well as to consider whether, and if so how, these limitations might eventually be overcome. As was the case for input heatmapping, the techniques of diagnostic classification and feature-detector-identification work relatively well when a system's variables can be interpreted semantically—that is, when they can be thought to represent recognizable features of the environment. Indeed, feature-detectors are only as informative as the features being detected can be recognized by human observers, and diagnostic classifiers require investigators to already possess a detailed understanding of the programs that *might* be executed by the system whose behavior is being explained. As has already been suggested, however, there are reasons to believe that the environmental features being detected and the correlations being learned through the use of Machine Learning may be subtle and difficult to interpret. For this reason, although promising, these XAI techniques are likely to be limited in scope and utility.

To what extent can this scope and utility be increased? Here it may once again be worth seeking guidance in cognitive science. There, one long-term trend is the gradual recognition that semantic interpretability is not always necessary to explain the behavior of humans and other biological cognizers. Indeed, it is now far less widely assumed than before that the neuronal processes described by cognitive models should—or even can—be described in semantic terms (Chemero, 2000; Ramsey, 1997). For this reason, many cognitive models today instead deploy sophisticated mathematical concepts and analytic techniques for idealizing, approximating, or otherwise reducing the dimensionality of the systems being investigated, even if they do not thereby render these systems semantically interpretable. Although the use of these concepts and techniques may render folk psychological categories inapplicable, they might nevertheless be used to satisfy important explanatory norms such as description, prediction, and intervention.

In conclusion, it seems fair to wonder whether Explainable AI might similarly move away from semantic interpretability, and toward idealization, approximation, and dimension-reduction. On the one hand, Artificial Intelligence is in part an engineering discipline tasked with developing technologies that improve the daily lives of regular individuals. For this reason, it is subject to constraints that pure sciences are not. Whereas it may not be important for laypeople to understand the processes that underlie visual perception, it is important that they know why their loan applications are getting rejected. On the other hand, just as society trusts scientists to possess expert knowledge that remains inaccessible to laypeople, it may accept that certain ML applications can only be explained by mastering the most sophisticated XAI techniques. In this case, it may be necessary to rely on societal mechanisms—for example, the testimony of an expert witness in a court of law—to ensure that an individual's rights are being protected even when that individual cannot him- or herself come to know the relevant facts. Thus, although semantic interpretability may often be beneficial, it may not always be essential.